\documentclass{ws-procs975x65}

\begin{document}

\title{Formation of Super-massive Black Holes}

\author{Wolfgang J. DUSCHL}

\address{Institut f\"ur Theoretische Astrophysik,\\
Tiergartenstr. 15,\\
69121 Heidelberg, Germany\\
E-mail: wjd@ita.uni-heidelberg.de}

\author{Peter A. STRITTMATTER}

\address{Steward Observatory, The University of Arizona,\\
933 N.\ Cherry Ave.,\\
Tucson, AZ 85721, USA\\
E-mail: pstrittmatter@as.arizona.edu}

\maketitle

\abstracts{ We show that the rapid formation of super-massive
black holes in quasars can indeed be understood in terms of major
galaxy mergers followed by disk accretion. The necessary short
disk evolution time can be achieved provided the disk viscosity is
sufficiently large, which, for instance, is the case for
hydrodynamic turbulence, unlimited by shock dissipation. We
present numerical calculations for a representative case. This
general picture can account for {\it (a)\/} the presence of highly
luminous quasars at redshifts $z > 6$; {\it (b)\/} for the peak in
quasar activity at $z \sim 2$; and {\it (c)\/} for a subsequent
rapid disappearance of quasars at later epochs.}

\section{Introduction}

From their observed redshift distribution, luminous quasars are
most prevalent at redshifts around $z \sim 2$
(Hasinger\cite{Has98}; Fan et al.\cite{FNL01}). Recent discoveries
have pushed back the limit at which galaxies and quasars appear in
the young Universe to redshifts of $z \sim 6.6$ for galaxies (Hu
et al.\cite{HCM02}) and $z \sim 6.4$ for quasars (Fan et
al.\cite{FSS03}, Willott et al.\cite{WMJ03}). These objects were,
therefore, already present when the Universe was less than $\sim
10^9$ years old\footnote{We take the following set of cosmological
parameters: $H_0 = 70\,$km\,s$^{-1}$\,Mpc$^{-1}$,
$\Omega_\mathrm{m} = 0.3$, $\Omega_\Lambda = 0.7$,
$\Omega_\mathrm{tot} = 1$, and a corresponding age of the Universe
of 13.5 Gyr}. Assuming that quasars are powered by accretion onto
super-massive black holes (SMBH) at rates at or below the
Eddington limit, luminosity measures from the Sloan Digital Sky
Survey (Fan et al.\cite{FNL01}) and from the Chandra and
XMM-Newton observatories (Brandt et al.\cite{BSF02,BVF02}), as
well as IR spectroscopy (Willott et al.\cite{WMJ03}) require that
black holes of mass $\ge 10^9\,\mathrm{M}_\odot$ are already
present at this early epoch. This leads to the question of the
origin of such SMBHs and, in particular, whether there is a viable
way of forming them in the very short time scale permitted by the
observational data.

It has been argued that the presence of massive black holes can be
understood in the framework of hierarchical merging, (e.g., Haiman
\& Loeb\cite{HLo01}). We suggest the accretion disk model presents
a viable alternative to the formation of super-massive black holes
and retains the advantage that it also provides a natural
explanation for the formation of jet outflows observed in many
quasars.

In this contribution, we will argue that the quasar phenomenon is
indeed a direct consequence of a {\it major merger\/} of
(proto-)galaxies followed by high-efficiency disk accretion onto a
black hole (c.f. Dopita\cite{Dop97} and references therein). We
define such a merger to be the coalescence of two gas-rich
galaxies of about equal mass resulting in the deposition of large
amounts of gas in a disk close to the center of the merged galaxy
(Barnes \& Hernquist\cite{BHe96,BHe98}, Naab \&
Burkert\cite{NBu01}, Barnes\cite{Bar02}). We then examine the
evolution of the disk through accretion driven by hydrodynamic
turbulence unlimited by shock dissipation (Duschl, Strittmatter \&
Biermann\cite{DSB00}, hereafter DSB) and show that the growth of a
central black hole can occur in the requisite short time scale. We
demonstrate that, even in the absence of massive BHs in the {\it
merging\/} galaxies, it is possible to form a sufficiently massive
BH in the {\it merged\/} galaxy in the required short time. While
the importance of mergers for feeding quasars has been under
discussion for some time (e.g., Stockton\cite{Sto99}, Canalizo \&
Stockton\cite{CSt01}), we will show that major mergers can be
instrumental both in {\it providing the fuel\/} and in {\it
building the engine\/} that produces the quasar phenomenon. The
general model also accounts for the absence of quasars at the
current epoch.

\section{The physical scenario}

In the following, we make the robust assumption that, due to a
major merger, tidal forces have driven a large amount
($10^{9\dots10}\,\mathrm{M}_\odot$) of accretable matter into the
central regions (within a few $10^2$ pc from the center) of the
newly formed merged galaxy. Detailed numerical model calculations
(Barnes \& Hernquist\cite{BHe96,BHe98}, Naab \&
Burkert\cite{NBu01}, Barnes\cite{Bar02}) have shown that in such
mergers, {\it (a)\/} the ISM loses most of its angular momentum
relatively rapidly, approximately on the dynamical timescale of
the galaxies involved, but {\it (b)\/} still retains too much
angular momentum to be immediately available for formation of or
accretion into a black hole. We assume that there is no
preexisting super-massive black hole at the center of the merged
galaxies, though we allow on numerical grounds for a comparatively
small seed black hole. This scenario provides the starting point
for our model.

We envisage that this self-gravitating disk of gas (and dust) will
evolve as follows. First material will accrete towards the center,
whether or not a seed (low mass) black hole is present, and will
be able to radiate all energy liberated through viscous
dissipation. The significant mass flow towards the disk's center
(see next section for details) will lead to {\it (a)\/} the
formation of a seed black hole (if none was present before), and
{\it (b)\/} an initial phase of Eddington-limited accretion into
it. We assume that the black hole accretes at its Eddington rate
as long as the disk delivers enough mass to maintain this
rate\footnote{We note that the strength of the Eddington limit on
the accretion rate is still not settled (e.g., Collin et
al.\cite{CBM02}, Ohsuga et al.\cite{OMM02}). Super-Eddington
accretion, however, is not required in the present model to
achieve the necessary time scales.}.

Ongoing accretion will deplete the mass of the accretion disk and
thus decrease the mass delivery rate towards the black hole, while
-- at the same time -- the black hole is growing in mass due to
the same accretion process. Ultimately the mass flow rate from the
disk to the black hole will become smaller than the Eddington
accretion rate,  {\it free accretion\/} will set in, and all
incoming mass will be accreted by the black hole. At this stage,
the accretion disk is still able to radiate all the energy
liberated by viscous dissipation. In the course of this evolution,
however, the accretion rate drops, both in absolute terms as well
as in units of the corresponding Eddington accretion rate. When
the actual accretion rate falls below roughly 0.3 \% of the
Eddington rate, the flow becomes advection dominated (Beckert \&
Duschl\cite{BDu02}), and the radiation efficiency of the accretion
process falls very quickly by several orders of magnitude. The
luminosity decreases correspondingly.

While the above scenario seems plausible for quasar evolution, the
question is whether it provides a quantitative explanation for the
observational data cited above. Earlier models (e.g., Shlosman,
Begelman, \& Frank\cite{SBF90}) based on $\alpha$-accretion disk
models (Shakura \& Sunyaev\cite{SSu73}) led to excessively long
evolution time scales (exceeding the Hubble time) for disks in the
centers of AGN, thereby precluding the formation of SMBHs at early
enough epochs. Consequently, various -- mostly non-axisymmetric --
processes (bars, spiral waves, etc.) were investigated in order to
speed up the accretion process (e.g., Shlosman, Frank, \&
Begelman\cite{SFB89}; Chakrabarti \& Wiita\cite{CWi93}), even
though disk models, because of their symmetry, provided a natural
origin for the collimated jet outflows, which appear to be a
frequent occurrence in quasars.

In the meantime DSB pointed out that experimental data on rotating
fluids suggest an alternative, hydrodynamic origin of turbulence
and hence a different viscosity prescription -- $\beta$-viscosity
-- that would apply as long as the associated turbulent motions
remained sub-sonic (see also Richard \& Zahn\cite{RZa99},
Longaretti\cite{Lon02}, and Richard\cite{Ric03}). In the following
section we investigate the evolution of accretion disks with such
$\beta$-viscosity.

\section{The model}

We have carried out numerical calculations that model the
evolution of an accretion disk resulting from a major merger. In
addition to the standard set of time dependent accretion disk
equations (see, e.g., Frank, King, \& Raine\cite{FKR02}), we take
self-gravity into account through Poisson's equation. The thermal
properties of the disk are treated with a single zone
approximation (for a discussion of the validity of this
approximation see Hur\'e \& Galliano\cite{HGa01}). We assume the
disk to be azimuthally symmetric and geometrically relatively thin
perpendicular to its rotational plane. For the initial
distribution of the mass in the disk, we chose a radial
distribution of the surface density $\Sigma \propto s^{-1}$ where
$s$ is the radial coordinate. If the mass flow rate to the black
hole exceeds the classical Eddington limit, we allow the black
hole to grow only at the corresponding Eddington rate.

For our models, we use the $\beta$-viscosity parameterization
suggested by DSB

\begin{equation}
\nu = \beta s v_\varphi
\end{equation}
whenever the turbulent velocity $v_\mathrm{turb} \sim \beta^{1/2}
v_\varphi$ and local sound velocity $c_\mathrm{s}$ in the
resultant flow satisfy the condition that

\begin{equation}
v_\mathrm{turb} \le c_\mathrm{s}\ .
\end{equation}
Laboratory experiments suggest a value of $\beta$ in the range
$10^{-3} < \beta < 10^{-2}$ , where $\beta^{-1}$ corresponds to
the critical Reynolds number $\Re_\mathrm{c}$ for the onset of
turbulence in the flow. The corresponding accretion and dynamical
time scales, $\tau_\mathrm{accr}$ and $\tau_\mathrm{dyn}$, are
given by

\begin{equation}
\tau_\mathrm{accr} = s^2/\nu = \left( \beta \omega \right)^{-1} =
\beta^{-1} \tau_\mathrm{dyn} = \Re_\mathrm{c} \tau_\mathrm{dyn}\ .
\end{equation}

In this framework it can be shown (Duschl \& Strittmatter, in
prep.) that the thermal timescale is sufficiently long compared to
the dynamical timescale to ensure stability of the disk against
fragmentation (Gammie\cite{Gam01}, Rice et al.\cite{RAB03}).

We use an explicit finite-difference scheme. At the inner radius
($s = s_\mathrm{i}$), we allow any material either to be accreted
onto the central black hole (at or below the Eddington rate) or to
be lost from the system. We also set the surface density $\Sigma
\left( s = s_\mathrm{i} \right) = 0$, or equivalently the viscous
torque $G \left( s_\mathrm{i} \right) = 0$. At the outer boundary
of the disk, we assume angular momentum to be removed efficiently.
We choose a fixed outer boundary at $s = s_\mathrm{o}$ and set
$\left. \partial G / \partial s \right|_{s = s_\mathrm{o}} = 0$,
which removes angular momentum from the material at the required
rate.

\begin{figure}
\epsfxsize=\textwidth
\epsfbox{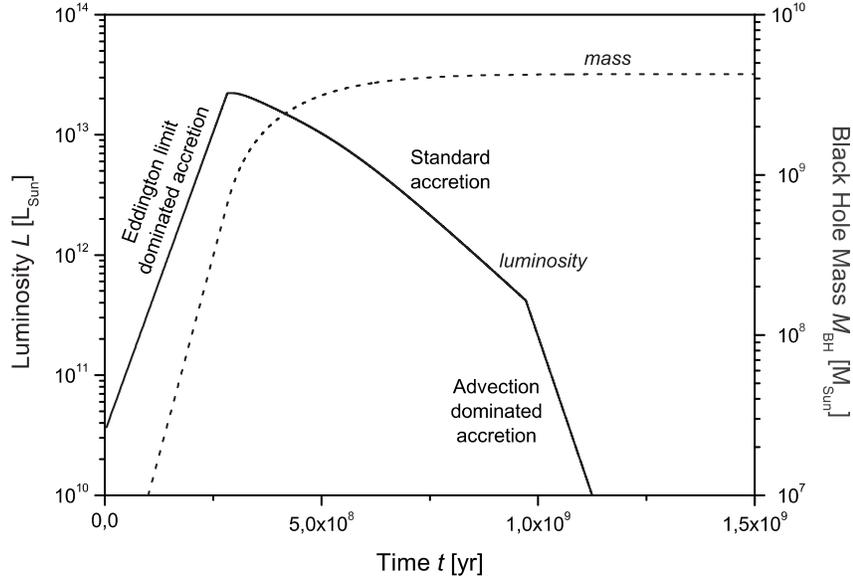} \caption{The evolution of
the accretion luminosity (full line) and of the mass (dotted line)
of a black hole in the center of an accretion disk with the
initial parameters given in the text. \label{fig:evolution}}
\end{figure}

As a specific illustrating example, we consider the evolution of
an accretion disk of initial mass $M_\mathrm{d} \left( t = 0
\right) = 10^{10}\,\mathrm{M}_\odot$, inner radius $s_\mathrm{i} =
10^{16}\,$cm, outer radius $s_\mathrm{o} = 10^{20}\,$cm, and
viscosity parameter $\beta = 10^{-3}$, and follow the evolution of
the mass of a central black hole. In the numerical model presented
here, we have assumed, for convenience, the initial presence of a
seed black hole mass of $10^6\,\mathrm{M}_\odot$. We have,
however, calculated models with seed black holes between $10^2$
and $10^7\,\mathrm{M}_\odot$, which lead qualitatively to the same
results.

The resulting evolution of quasar luminosity and black hole mass
are illustrated in Figure \ref{fig:evolution}. For the first
$3\cdot 10^8\,$ years, the growth of the black hole mass is
constrained by the Eddington limit. This time scale corresponds to
Salpeter's\cite{Sal64} growth time scale and as such depends on
the initial mass of the central object at $t = 0$.

After that time, the free accretion period sets in, during which
the central black hole is able to swallow all matter delivered by
the disk. However, due to the decreasing disk mass (still large
compared to the central black hole) the mass flow into the black
hole slows down and the luminosity declines correspondingly. At
the same time, due to an ever-increasing black hole mass, the
limiting Eddington rate continues to increase. In this example,
the free accretion phase lasts for $7\cdot 10^8\,$years. It comes
to an end when the accretion rate falls below 0.3\,\% of the
Eddington rate and the flow becomes advection dominated (Beckert
\& Duschl\cite{BDu02}). While the accretion rate itself continues
to fall slowly, the radiation efficiency of the accreting material
drops drastically, and consequently so does the accretion
luminosity. We note that the time spent at a luminosity exceeding
one half the peak value is roughly $2.5\cdot 10^8\,$years, mostly
in the post-peak era.

\section{Discussion}

Our results (Fig.\ \ref{fig:evolution}) show that the luminous ($>
10^{11}\,\mathrm{L}_\odot$), and hence readily detectable, phase
of our proposed quasar model lasts $\le 10^9\,$years. While the
precise duration of the black hole accretion phase will depend on
the detailed parameters of the disk formed during the major
merger, a typical timescale from the onset of the disk evolution
to the beginning of the advection dominated phase of $\sim
10^9\,$years should be representative. The duration of the most
luminous phase around the transition from the Eddington limited to
free accretion is much shorter than $10^9\,$years, in qualitative
agreement with recent observational results (e.g., Yu \&
Tremaine\cite{YTr02}). This is significantly shorter than the
interval ($\sim 10^{10}\,$years corresponding to redshifts $1 < z
< 5$) during which quasars are prevalent in the Universe. Within
this model, therefore, the number density of quasars at different
epochs is determined almost entirely by the rate of occurrence of
major mergers. This question has been analyzed by several authors
(e.g., Kauffmann \& Haehnelt\cite{KHa00}, Duschl \& Horst, in
prep.) and has been shown to peak at epochs corresponding to $z
\sim 2$, a result that is consistent with the observed
distribution of quasars (Hasinger\cite{Has98}). The quasar with
the currently highest known redshift, SDSS J1148+5251, at $z =
6.4$, seems to still be in the Eddington limit controlled regime
(Willott et al.\cite{WMJ03}, Barth et al.\cite{BMN03}), in all
likelihood not too far from its maximum luminosity. Taking all
this evidence together, the model, therefore, seems to be broadly
consistent with the available observational data.

For the above quasar scenario to succeed, it is essential to have
an efficient accretion process so that a black hole can grow
quickly and can produce the required luminosity. The accretion
process also has to be efficient enough to accrete away most of
the available gas and dust and thus lead to a rapid end of the
quasar phenomenon due to a drop in both the accretion rate and in
the disk's radiation efficiency. The time scales computed above
for the various phases of $\beta$-disk evolution are substantially
shorter than those previously derived for $\alpha$-disk models.

The disk's viscosity is therefore the crucial quantity. As pointed
out by DSB and by Richard and Zahn\cite{RZa99}, laboratory
experiments indicate that the $\beta$-prescription is appropriate
for turbulent viscosity in incompressible flows where the
turbulence is clearly driven hydrodynamically. DSB suggested that
this prescription is also appropriate in compressible flows, such
as accretion disks, provided the turbulence remains sub-sonic so
that shock dissipation is negligible\footnote{DSB also show that
in the case of shock limited turbulence (i.e., when the
hydrodynamically driven turbulence would be super-sonic) in
Keplerian disks, the $\alpha$-prescription is indeed appropriate.
This situation applies, for example, to the disks in cataclysmic
variable stars.}. For internal consistency this model, therefore,
requires that throughout the disk flow $\delta = v_\mathrm{turb} /
c_\mathrm{s} = \beta^{1/2} v_\varphi / c_\mathrm{s} \le 1$. In the
calculations reported above, the ratio $\delta$ satisfies the
condition $0.01 \le \delta \le 1$ throughout the disk and at all
times, so that the model remains internally self consistent. As
noted above this ``hot disk" model also carries with it the
consequence that the disk, while flattened, cannot be very thin
and will be stable against fragmentation (Duschl \& Strittmatter,
in prep.).

We acknowledge that angular momentum can be removed rapidly from
the disk through other mechanisms -- usually involving
non-axisymmetric instabilities -- so that the $\beta$-disk model
is not unique in providing rapid time scales. On the other hand,
because of their symmetry, disk models do provide a natural
scenario for the generation of collimated jet outflows, which
appear to be a frequent occurrence in quasars. It is noteworthy
that while among local AGNs of the Seyfert type a larger fraction
of the host galaxies may be barred than among non-active but
otherwise similar galaxies, this fraction is clearly below unity
(Laine et al.\cite{LSK02}). This renders bar action as the (sole)
driving mechanism of AGNs highly unlikely.

The model described above has been highly simplified, in that we
have not treated star formation, mass-loss from the disk itself or
the fate of matter that could not be assimilated by the black hole
during the Eddington limited phase of disk evolution. In reality,
a significant part of the material in a self-gravitating disk will
be transformed into stars. However, the matter supply from the
disk is more than sufficient to maintain an Eddington accretion
rate for several $10^8\,$years. The essential features of the
model will thus remain unless virtually all ($> 90\ $\%) of the
disk material is transformed into stars. We will address the role
of star formation in self-gravitating accretion disks in an
upcoming paper (Duschl \& Strittmatter, in prep.). In regard to
mass loss from the disk, especially near the black hole, one may
speculate that this provides an ideal source of material and
energy to form a jet and a broad line region. It is also possible
that the later (advection dominated phase) in which the thermal
luminosity is small, may result in increased visibility of
non-thermal jet emission and hence the blazar phenomenon towards
the end of the quasar lifetime.

Given the short formation time scale for massive black holes, the
present scenario obviates the need to postulate the existence of
primordial SMBH in accounting for the quasar phenomenon. The model
requires that quasars occur in galaxies which encountered major
mergers so that in today's Universe, these galaxies must have
massive ($10^9\ \mathrm{M}_\odot$ or more) central black holes.
Galaxies, which never experienced a major merger, may harbor black
holes of considerably smaller mass and may, therefore, still
exhibit phases of more modest nuclear activity, for instance as
Seyfert galaxies. Clearly as the strength of galaxy interactions
varies so also will the observable characteristics of the merged
galaxy (or post interaction galaxies). For example less accretable
mass driven into a more extensive disk, would make the
corresponding time scales of viscous evolution much longer and the
central source less luminous. Such sources may well be associated
with the faint, optically selected AGN population noted by Steidel
et al.\cite{SHS02}.

There is mounting evidence for a close relation between the mass
of a central black hole, the bulge velocity dispersion (Ferrarese
\& Merritt\cite{FMe00}; Gebhardt et al.\cite{GRK00}), and the
galaxy's circular velocity (Ferrarese\cite{Fer02}).  The proposed
scenario does not, in itself, naturally predict such an effect,
although it does not exclude it either.

\section{Summary}

We have discussed a model for the origin of quasars in which a
major merger of galaxies results in the creation first of a
central self-gravitating accretion disk. In such an environment,
under the influence of hydrodynamically induced $\beta$-viscosity,
the disk evolves much more rapidly than predicted by standard
($\alpha$-)disk theory. This evolution leads through three stages,
namely first Eddington-limited, then free, and finally advection
dominated accretion. The model seems capable of explaining the
early epoch of the first quasars, their epoch of peak activity at
redshifts around $z \sim 2$ and their subsequent rapid
disappearance. The proposed scenario provides not only the fuel
for the quasar phenomenon but also the creation of the SMBH
engine.

\section*{Acknowledgments}

WJD acknowledges generous support from Steward Observatory and
from the Deutsche Forschungsgemeinschaft through SFB 439.


\begin{thebibliography}{99}

\bibitem{Bar02}Barnes, J.E., {\it MNRAS\/} {\bf 333}, 481 (2002)

\bibitem{BHe96}Barnes, J.E., Hernquist, L., {\it ApJ\/} {\bf 471},
115 (1996)

\bibitem{BHe98}Barnes, J.E., Hernquist, L., {\it ApJ\/} {\bf 495},
187 (1998)

\bibitem{BMN03}Barth, A.J., Martini, P.,, Nelson, C., Ho,
L.C., {\it ApJ\/} {\bf 594}, L95 (2003)

\bibitem{BDu02}Beckert, T., Duschl, W.J., {\it A\&A} {\bf 387}, 422
(2002)

\bibitem{BSF02}Brandt, W.N., Schneider, D.P., Fan, X., et al.,
{\it ApJ\/} {\bf 569}, L5 (2002)

\bibitem{BVF02}Brandt, W.N., Vignali, C., Fan, X., et al.,
{\it MPE-Report\/} {\bf 279}, 235 (2002)

\bibitem{CSt01}Canalizo, G., Stockton, A., {\it ApJ\/} {\bf 555},
719 (2001)

\bibitem{CWi93}Chakrabarti, S.K., Wiita, P.J., {\it ApJ\/} {\bf 411},
602 (1993)

\bibitem{CBM02}Collin, S., Boisson, C., Mouchet, M., et al., {\it
A\&A\/} {\bf 388}, 771 (2002)

\bibitem{Dop97}Dopita, M.A., {\it PASA\/} {\bf 14}, 230 (1997)

\bibitem{DSB00}Duschl, W.J., Strittmatter, P.A., Biermann, P.L., {\it
A\&A\/} {\bf 357}, 1123 (2000) (=DSB)

\bibitem{FNL01}Fan, X., Narayanan, V.K., Lupton, R.H., et al.,
{\it AJ\/} {\bf 122}, 2833 (2001)

\bibitem{FSS03}Fan, X., Strauss, M.A., Schneider, D.P., et al., {\it
AJ\/} {\bf 125}, 1649 (2003)

\bibitem{Fer02}Ferrarese, L., {\it ApJ\/} {\bf 578}, 90 (2002)

\bibitem{FMe00}Ferrarese, L., Merritt, D., {\it ApJ\/} {\bf 539}, L9
(2000)

\bibitem{FKR02}Frank, J., King, A., Raine, D., {\it Accretion Power
in Astrophysics\/} (3rd ed.), Cambridge University Press,
Cambridge, UK (2002)

\bibitem{Gam01}Gammie, C.F., {\it ApJ\/} {\bf 553}, 174 (2001)

\bibitem{GRK00}Gebhardt, K., Richstone, D., Kormendy, J., et al.,
{\it AJ\/} {\bf 119}, 1157 (2000)

\bibitem{HLo01}Haiman, Z., Loeb, A., {\it ApJ\/} {\bf 552}, 459
(2001)

\bibitem{Has98}Hasinger, G., {\it AN\/} {\bf 319}, 37 (1998)

\bibitem{HCM02}Hu, E.M., Cowie, L.L., McMahon, R.G., {\it ApJ\/}
{\bf 568}, L75 (2002)

\bibitem{HGa01}Hur\'e, J.-M., Galliano, F., {\it A\&A\/} {\bf
366}, 359 (2001)

\bibitem{KHa00}Kauffmann, G., Haehnelt, M., {\it MNRAS\/} {\bf 311},
576 (2000)

\bibitem{LSK02}Laine, S., Shlosman, I., Knapen, J., Peletier, R.F.,
{\it ApJ\/} {\bf 567}, 97 (2002)

\bibitem{Lon02}Longaretti, P.-Y., {\it ApJ\/} {\bf 576}, 587 (2002)

\bibitem{NBu01}Naab, Th., Burkert, A., {\it ASP Conf. Ser.\/}{\bf
249}, 735 (2001)

\bibitem{OMM02}Ohsuga, K., Mineshige, S., Mori, M., Umemura M.,
{\it ApJ\/} {\bf 574}, 315 (2002)

\bibitem{RAB03}Rice, W.K.M., Armitage, P.J., Bate, M.R., Bonnell,
I.A., {\it MNRAS\/} {\bf 339}, 1025 (2003)

\bibitem{Ric03}Richard, D.T., {\it A\&A\/} {\bf 408}, 409 (2003)

\bibitem{RZa99}Richard, D., Zahn, J.-P., {\it A\&A\/} {\bf 347},
734 (1999)

\bibitem{Sal64}Salpeter, E.P., {\it ApJ\/} {\bf 140}, 796 (1964)

\bibitem{SSu73}Shakura, N.I., Sunyaev, R.A., {\it A\&A\/} {\bf
24}, 337 (1973)

\bibitem{SBF90}Shlosman, I., Begelman, M.C., Frank, J., {\it
Nature\/} {\bf 345}, 679 (1990)

\bibitem{SFB89}Shlosman, I., Frank, J., Begelman, M.C.,
{\it Nature\/} {\bf 338}, 45 (1989)

\bibitem{SHS02}Steidel, C.C., Hunt, M.P., Shapley, A.E., {\it
ApJ\/} {\bf 576}, 653 (2002)

\bibitem{Sto99}Stockton, A., {\it IAU-Symp.\/} {\bf 186}, 311
(1999)

\bibitem{WMJ03}Willott, C.J., McLure, R.J., Jarvis, M.J.,
{\it ApJ\/} {\bf 587}, L15 (2003)

\bibitem{YTr02}Yu, Q.-J., Tremaine, S., {\it MNRAS\/} {\bf 335},
965 (2002)

\end{thebibliography}
\end{document}